%% file: main.tex
\documentclass[%
reprint,
superscriptaddress,
]
{revtex4-2}

\input{1packages}
\input{2commands}
\begin{document}
\input{3title}
\maketitle

\section{\label{sec:intro}INTRODUCTION\protect}

Entangled states are fundamental for the development of quantum information processing platforms. At the same time, their preparation is motivated by the urge to understand the nonlocality of nature. Traditionally, interactions with the environment, particularly those leading to dissipation, have been considered an obstacle and a source of decoherence for achieving entangled states~\cite{DiVincenzo_1998}. More recently, we have learned to use dissipation to our advantage by engineering the dynamics of certain systems to drive them into a desired entangled state~\cite{diehl_quantum_2008,Verstraete2009,Vacanti_2009,Shen_2011,Kastoryano2011,Reiter_2012,Stannigel_2012,Muschik_2011,DallaTorre_2013,Rao_2014,Lin_2013,krauter_entanglement_2011,MartinCano_2011,tudela_mesoscopic_2013,ballestero_chiral_2015,Gonzalez2015}. Different protocols have been proposed for platforms such as cavity quantum electrodynamics (QED)~\cite{Kastoryano2011,Shen_2011,Reiter_2012,Stannigel_2012}, atomic ensembles~\cite{Muschik_2011,DallaTorre_2013,Rao_2014,Lin_2013,krauter_entanglement_2011}, waveguide QED~\cite{MartinCano_2011,tudela_mesoscopic_2013,ballestero_chiral_2015,Gonzalez2015,Tiranov2023,Santos_2023}, trapped ions~\cite{barreiro_2011}, and superconducting qubits~\cite{reiter_2013,li_observation_2025,shankar_2013,brown_trade_2022}. These protocols generally drive the system into an entangled state independently of the starting configuration. Ultimately, the concept of utilizing dissipation for the preparation of entangled states can be extended to multiple tasks in quantum information processing, including universal quantum computation~\cite{Verstraete2009}. 

In this paper, we focus on the dissipative preparation of steady-state multiparticle entangled states in waveguide QED. We present a simple scalable protocol that can be implemented to generate high-fidelity W-type entangled states~\cite{dur_2000}, without requiring feedback control mechanisms or precise timing of pulses. Our system consists of multiple $\Lambda$-type emitters, and the desired state is an entangled state of the ground levels. Starting from an arbitrary state, the dynamics drive the system into an entangled state by using the presence of subradiant and superradiant states~\cite{van_loo_photon_2013,albrecht_subradiant_2019} for multiple emitters coupled to the waveguide. Subradiant and superradiant states can have markedly different decay rates for emitters well coupled to a waveguide. This phenomenon is combined with the quantum Zeno effect (QZE)~\cite{Itano_90}. Here, the QZE treats the environment as an effective measuring device of decaying states, with a measurement rate set by the decay rate. For a system that evolves slowly enough, transitions into states that are observed more frequently are suppressed. Consequently, a system initially in its ground state and coupled to both subradiant and superradiant states has a higher probability of transitioning through a subradiant state than through 
a superradiant one. We exploit these properties to engineer the dynamics, ensuring that the desired state is the only one for which the pumping rate into the state exceeds the pumping rate out of it.

We show that the steady-state fidelity $F$ grows with the coupling efficiency $\beta$~\cite{Sheremet2023}, and we find that it scales as $(1-F)\sim 1/C$, where $C\equiv\beta/(1-\beta)$ is the cooperativity. This scaling with $C$ is distinctive of dissipative preparation of states \cite{Kastoryano2011}. In contrast, in the creation of entangled states by the implementation of unitary gates, decay is a more severe source of errors, and the fidelity characteristically scales as $(1-F)\sim 1/\sqrt{C}$~\cite{Anders_2003,Pellizarri95,Domokos95,Borregaard2015}. Moreover, we show that our protocol allows for the feasible preparation of high-fidelity states for current state-of-the-art experimental platforms.

The article is structured as follows. First, in Sec.~\ref{sec:scheme}, we present the scheme and derive the final fidelity as the main figure of merit. In this section, we also discuss the intrinsic errors of the protocol. In Sec.~\ref{sec:scaling}, we generalize the scheme to a larger number of emitters. In Sec.~\ref{sec:example}, we provide a realistic implementation of the scheme for trapped $^{133}$Cs atoms. We explain how we envision reaching the strong-waveguide coupling regime and include some additional experimental errors such as broadening of the ground states and atomic motion. We present a possible alternative protocol that exploits microwave transitions in the Appendix.

\section{\label{sec:scheme}TWO-EMITTERS SCHEME}

\begin{figure}[t]
\centering
    \includegraphics{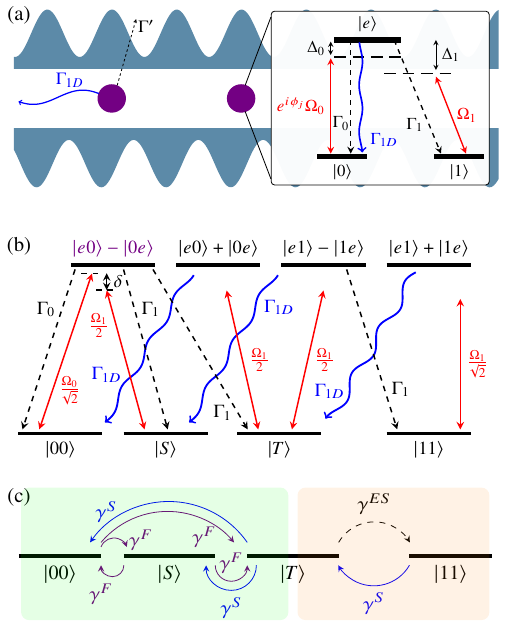}
\caption{\label{fig:1} (a) Schematic of the system. Two emitters are located in an optical waveguide. The inset shows the level structure of the emitters. The emitters decay to state $\ket{0}$ through the waveguide at an enhanced rate $\Gamma_\mathrm{1D}$ and to the side with decay rates $\Gamma_0$ and $\Gamma_1$ to states $\ket{0}$ and $\ket{1}$, respectively. $\Omega_i$ and $\Delta_i$ represent the strength and the detuning, respectively, for the driving between the ground state $\ket{i}$ and the excited state. The decay from $\ket{e}$ to $\ket{0}$ couples to the waveguide but not the one to $\ket{1}$. We off-resonantly drive the ground states to the excited ones. Note that we add a phase difference between the drivings on $\ket{0}_j$ to couple the state $\ket{00}$ to the subradiant manifold. (b) Representation of singly excited subradiant (purple) and superradiant (black) manifold, as well as the ground states, where $\ket{T}$ is the desired state. $\ket{00}$ and $\ket{S}$ couple to a subradiant state, resulting in rapid pumping into $\ket{T}$ at a rate $\sim\gamma^F$. $\ket{T}$ is pumped back through superradiant states, suppressing the process to a slower rate $\sim\gamma^S$. On the other hand, $\ket{T}$ effectively decays to $\ket{11}$ through nonguided decay from a superradiant state, at an extremely slow rate $\gamma^{ES}$. The decay from $\ket{11}$ to $\ket{T}$ occurs through superradiant decay at the rate $\gamma^S$. (c) Effective ground states evolution separated into fast (green box) and slow (orange box) dynamics.} 
\end{figure}

For simplicity, we first present the protocol for two emitters. We assume the emitters to have a $\Lambda$-type level structure, see Fig.~\hyperref[fig:1]{1(a)}. Each emitter has two ground states, $\ket{0}$ and $\ket{1}$, and an excited state, $\ket{e}$, which decays to both ground states. Our scheme requires one of the transitions to be well coupled to a bidirectional waveguide, which enhances that decay rate, i.e., the emitters are placed close to a waveguide, and the levels are chosen such that the transition $\ket{e}\leftrightarrow\ket{0}$ is enhanced by the coupling to the waveguide, but not the transition $\ket{e}\leftrightarrow\ket{1}$. Additionally, the protocol relies heavily on the distance between emitters to be an integer times the $\ket{e}\leftrightarrow\ket{0}$ transition wavelength in the waveguide. Thus, the emitters are coupled dissipatively and not dispersively~\cite{van_loo_photon_2013,albrecht_subradiant_2019}. As discussed in the following, this means that some excited states decay with superradiant decay at a rate of $\sim \Gamma_\mathrm{1D}+\Gamma'$, where $\Gamma_\mathrm{1D}$ is the decay rate into the waveguide, $\Gamma'=\Gamma_0+\Gamma_1$ is the total nonguided decay rate, and $\Gamma_n$ is the nonguided decay rate to the ground state $\ket{n}$. 
The rest of the excited states, which we call subradiant states,  decay at a rate $\Gamma'$. For two emitters, there is a single unique subradiant state, see Fig.~\hyperref[fig:1]{1(b)}. The presented protocol works for well-coupled emitters to the waveguide, meaning that we consider the regime $\Gamma_\mathrm{1D}\gg\Gamma'$.

Similar protocols limited to preparing two-particle entangled states, including those presented in Ref.~\cite{Reiter_2012} and the Appendix, consider a microwave drive between the ground states. The scalable protocol presented here works by employing a weak optical drive between the ground and excited states, $\Omega_n\ll\Gamma'$, where $\Omega_n$ is the driving strength between $\ket{e}$ and the ground state $\ket{n}$. Due to the fast decay of the excited states, this means that (1) the system will mostly remain within the ground-state manifold and, therefore, (2) we can adiabatically eliminate the excited states and study the system solely based on the ground states' evolution. For two emitters, the objective of the protocol is to generate a high-fidelity state with respect to the maximally entangled state
\begin{equation}
\ket{T}=\frac{1}{\sqrt{2}}(\ket{0}_1\ket{1}_2+\ket{1}_1\ket{0}_2),
\end{equation}
where $\ket{n}_j$ stands for the ground state $\ket{n}$ from the $j$th emitter.

To understand the protocol, we choose to study the system's evolution with the ground states given in the singlet-triplet basis: \{$\ket{00}=\ket{0}_1\ket{0}_2$, $\ket{T}=\frac{1}{\sqrt{2}}(\ket{0}_1\ket{1}_2+\ket{1}_1\ket{0}_2)$, $\ket{S}=\frac{1}{\sqrt{2}}(\ket{0}_1\ket{1}_2-\ket{1}_1\ket{0}_2)$, $\ket{11}=\ket{1}_1\ket{1}_2$\}. In Fig.~\hyperref[fig:1]{1(b)}, we depict the system and its dynamics, including the ground states and the singly-excited states. Due to the weak driving, the main contribution to a transition between two ground states results from the driving to an intermediate excited state and subsequent decay. We refer to these kinds of transitions as effective decays. Due to the QZE, excitations to intermediate excited states are suppressed when those states decay rapidly compared to states mediated by more slowly decaying ones. Consequently, transitions mediated by superradiant states have smaller effective decay rates than those mediated by subradiant states. As shown in Fig.~\hyperref[fig:1]{1(b)}, we engineer the evolution such that both $\ket{00}$ and $\ket{S}$ are coupled to the subradiant state, which subsequently decays to $\ket{T}$, while $\ket{T}$ and $\ket{11}$ couple exclusively to superradiant states. This is achieved by introducing a phase difference between the drivings on $\ket{0}_j$, which couples the state $\ket{00}$ to the subradiant manifold. As a result, there is fast pumping through the subradiant states from $\ket{00}$ and $\ket{S}$ into $\ket{T}$, whereas these states can be repumped from $\ket{T}$ only via superradiant states. In the effective dynamics shown in Fig.~\hyperref[fig:1]{1(c)}, $\ket{00}$ and $\ket{S}$ are quickly pumped to $\ket{T}$ at a fast rate $\sim\gamma^F$, while $\ket{T}$ decays back at a much slower rate $\sim\gamma^S$. In contrast, although $\ket{11}$ couples exclusively to superradiant states, only nonguided decay channels from the superradiant manifold lead the system to $\ket{11}$. As a consequence, this process happens at a rate $\sim\gamma^{ES}$ that is even slower than $\sim\gamma^S$. There is therefore very little population in $\ket{11}$ and almost all the population accumulates in  $\ket{T}$. Further details are provided in the next section.

\subsection{DYNAMICS\label{sec:dynamics}}

We now perform a detailed analysis of the dynamics of the system. The Hamiltonian of the system is given by 
\begin{equation}
H = H_0+V_++V_-,
\label{eq:ham1.1}
\end{equation}
\begin{equation}
H_0 = -\sum_{j=1,2}\left[\Delta_0\ketbra{0}{0}_j+\Delta_1\ketbra{1}{1}_j\right],
\label{eq:ham1.2}
\end{equation}
\begin{equation}
V_+ = \sum_{j=1,2}\left[\frac{\Omega_0}{2}e^{i\phi_j}\ketbra{e}{0}_j+\frac{\Omega_1}{2}\ketbra{e}{1}_j\right],
\label{eq:ham1.3}
\end{equation}
where $V_-=V_+^\dagger$, and $\ket{0}_j$ and $\ket{1}_j$ are the $j$th atom ground states for $j\in\{1,2\}$. We drive the ground states to the excited state $\ket{e}_j$ with Rabi frequencies $\Omega_0$, and $\Omega_1$ and detunings $\Delta_0$ and $\Delta_1$, respectively. $V$ represents the coupling between ground- and excited-state manifolds. The phase on the driving from $\ket{0}_j$ is ideally given by $e^{i\phi_j} = (-1)^j$. On the other hand, the nonunitary evolution is given by the Lindblad operators
\begin{equation}
L_{j}^{(0)} = \sqrt{\Gamma_0}\ketbra{0}{e}_j,
\label{eq:lind1.1}
\end{equation}
\begin{equation}
L_{j}^{(1)} = \sqrt{\Gamma_1}\ketbra{1}{e}_j,
\label{eq:lind1.2}
\end{equation}
\begin{equation}
L^{(c)} = \sqrt{\Gamma_\mathrm{1D}}\left[\ketbra{0}{e}_1+\ketbra{0}{e}_2\right],
\label{eq:lind1.3}
\end{equation}
where $L_{j}^{(0)}$ and $L_{j}^{(1)}$ correspond to nonguided decay processes and $L^{(c)}$ to a collective one. The latter arises from the emitters being dissipatively coupled through the waveguide. Note that here we assume that the coupling between emitters is exclusively dissipative, and there is no dispersive interaction between them~\cite{Tiranov2023}. We also assume the waveguide coupling efficiency to be close to unity, which implies $\Gamma_\mathrm{1D}\gg\Gamma_0,\Gamma_1$. 

From these Lindblad operators, we identify the excited-state manifold as the set of states susceptible to decay. Importantly, this manifold can be divided into states that decay via collective (superradiant) processes and are affected by $L^{(c)}$, and states that undergo only nonguided (subradiant) decays; see Fig.~\hyperref[fig:1]{1(b)}.

We are motivated to use the effective operator formalism~\cite{ReiterSorensen2012}, since we consider a weak drive between the ground and excited states. This approach allows us to adiabatically eliminate the rapidly evolving states---here, the excited states---and focus on the dynamics of the more slowly evolving ground states. In the effective operator formalism, the effective Hamiltonian $H_{\mathrm{eff}}$ and the effective Lindblad operators $L_{\mathrm{eff}}^{(k)}$ are given by
\begin{equation}
H_{\mathrm{eff}}=-\frac{1}{2}\left[V_-\sum_{n,j} \left(H_\mathrm{NH}^{(n,j)}\right)^{-1}V_+^{(n,j)}+\mathrm{H.c.}\right]+H_g,
\end{equation}
\begin{equation}
L_{\mathrm{eff}}^{(k)}=L^{(k)}\sum_{n,j} \left(H_\mathrm{NH}^{(n,j)}\right)^{-1}V^{(n,j)}_+,
\end{equation}
where we refer to an arbitrary Lindblad operator as $L^{(k)}$. The ground-state Hamiltonian in the singlet–triplet basis is given by
$H_g=-2\Delta_0\ketbra{00}{00}-2\Delta_1\ketbra{11}{11}-(\Delta_0+\Delta_1)(\ketbra{T}{T}+\ketbra{S}{S})$. For each transition from the ground state $\ket{n}_j$ to an excited state, we compute the quantity $\left(H_\mathrm{NH}^{(n,j)}\right)^{-1} V_+^{(n,j)}$, where, for this scheme, it is given by
\begin{equation*}
H_\mathrm{NH}^{(n,j)}=\Delta_n-\frac{i}{2}\sum_{k} L^{(k)\dagger} L^{(k)}
\end{equation*}
since $\Delta_n$ does not depend on the emitter. $H_\mathrm{NH}^{(n,j)}$ is the transition-dependent non-Hermitian Hamiltonian, and $V_+^{(n,j)}$ denotes the matrix element defined in Eq.~(\ref{eq:ham1.3}) corresponding to a transition from $\ket{n}_j$.

We find that the singlet-triplet basis provides a clear picture of the dynamics; see Fig.~\hyperref[fig:1]{1(b)}. Moreover, this basis is advantageous for computing the steady state, since its states are eigenstates of $H_g$. From the dynamics given in Eqs.~(\hyperref[eq:ham1.1]{2})-(\hyperref[eq:lind1.3]{7}), the effective Lindblad operators in this basis read
\begin{equation}
L_{\mathrm{eff},j}^{(0)}=\sqrt{\gamma^{F}_{00}}\ketbra{00}{00}+\sqrt{{\gamma^{F}_{01}}/{2}}\ketbra{00}{S},
\label{eq:eff1.1}
\end{equation}
\begin{align}
L_{\mathrm{eff},j}^{(1)}=&\sqrt{\gamma^{F}_{10}/2}\,(\ket{T}+(-1)^j\ket{S})\bra{00}\nonumber\\&+\sqrt{\gamma^{F}_{11}/4}\,(\ket{T}+(-1)^j\ket{S})\bra{S}+\sqrt{\gamma^{ES}_{10}/2}\ketbra{11}{T},\quad\quad
\label{eq:eff1.2}
\end{align}
\begin{equation}
L_{\mathrm{eff}}^{(c)}=\sqrt{\gamma^{S}_{1}/2}\ketbra{00}{T}+\sqrt{\gamma^{S}_{0}}\ketbra{S}{T}+\sqrt{2\gamma^{S}_{1}}\ketbra{T}{11},
\label{eq:eff1.3}
\end{equation}
where the different effective decay rates are given by
\begin{equation*}
\gamma^{F}_{nm}=\frac{\Gamma_n\Omega_m^2}{4\Delta_m^2+\Gamma'^2},
\label{eq:gammaF}
\end{equation*}
\begin{equation*}
\gamma^{S}_{m}=\frac{\Omega_m^2}{\Gamma_\mathrm{1D}+\Gamma'},
\label{eq:gammaS}
\end{equation*}
and
\begin{equation*}
\gamma^{ES}_{nm}=\frac{\Gamma_n\Omega_m^2}{(\Gamma_\mathrm{1D}+\Gamma')^2},
\label{eq:gammaES}
\end{equation*}
where $n,m\in\{0,1\}$ and such that $\gamma^{F}_{nm}\gg\gamma^{S}_{m}\gg\gamma^{ES}_{nm}$ for emitters well coupled to the waveguide. From these effective decay rates, we can observe the consequences of the quantum Zeno effect: (1) Transitions that are mediated by the subradiant state are proportional to the fast effective decay rate $\gamma^{F}_{nm}$, which is large because of the small denominator compared to the other terms ($\Delta_j,\Gamma'\ll\Gamma_\mathrm{1D}$), and (2) transitions mediated by superradiant states are proportional to the slow effective decay rate $\gamma^{S}_{m}$, or to the even slower rate $\gamma^{ES}_{nm}$. These are small because of the large denominator, and in the latter case further suppressed by the small numerator. 

\begin{figure}[ht!]
\centering
    \includegraphics{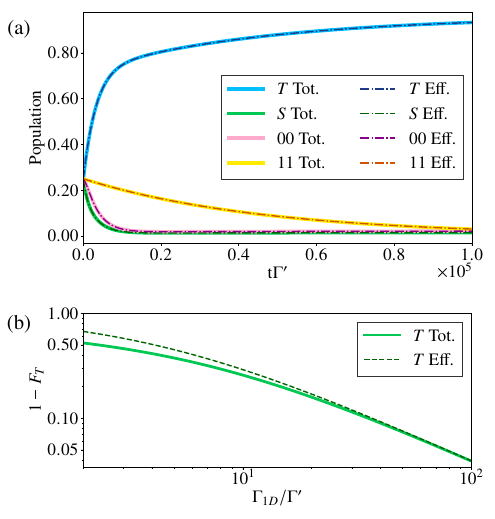}
\caption{\label{fig:2} (a) Comparison of the transient evolution of the system obtained from the total system dynamics (solid) and from the effective dynamics restricted to ground states (dashed). Starting from a completely mixed state, we observe excellent agreement between the two approaches, as well as how the population of the system converges into the entangled state $\ket{T}$. We have used $\beta=0.99$. (b) Fidelity as a function of the cooperativity $C=\Gamma_{1D}/\Gamma'$, comparing the total system numerical result (solid) with the analytical prediction from Eq.~(\ref{eq:errstat}). We observe good agreement between them, in particular, for large $C$. For both figures, we use $\Gamma_1=\Gamma_2$, $\Omega_0=\Delta_0=-\Delta_1=\Gamma'/20$, and the optimal driving conditions $\Omega_1^2=\Omega_0^2\mathcal{R}$; see Eqs.~(\ref{eq:optim_ratio},\ref{eq:opt_r}).}
\end{figure}

In Fig.~\hyperref[fig:2]{2(a)}, we compare the total system dynamics, found numerically, considering all the states of the system, with the effective dynamics between the ground states and show a perfect match between them. We also see how these dynamics drive the population of the system into the entangled-state $\ket{T}$ that we are aiming for.

In Eqs.~(\ref{eq:eff1.1}-\ref{eq:eff1.3}), we have only included the detunings $\Delta_i$ for the effective decays mediated by the subradiant state, since they are the most relevant for those processes. The detunings reduce the effective decay rate for pumping into $\ket{T}$, suggesting that minimizing them improves the accumulation of population in the desired state. However, as we will see in Sec.~\ref{sec:intrinsicerrors}, some detuning is needed to avoid stagnation in a dark state.

In the weak coupling regime $H_{\mathrm{eff}}\approx H_g$, such that $H_g$ is diagonal in the singlet-triplet basis. The errors resulting from this approximation are analyzed in the following section. Under this approximation, we neglect the off-diagonal elements giving coherences between the different ground states, and it allows us to describe the population dynamics using simplified rate equations. In these equations, the effective decay rates determine the exchange rates between the ground states~\cite{Reiter_2012}. From them, we find that the steady-state fidelity of $\ket{T}$, $F_{T}=\bra{T}\rho\ket{T}$, where $\rho$ is the density matrix, is given by
\begin{align}
\frac{1-F_{T}}{F_{T}}=&\frac{\Gamma_1}{2(\Gamma_\mathrm{1D}+\Gamma')}\frac{\Omega_0^2}{\Omega_1^2}+\frac{\Gamma'^2+4\Delta_1^2}{2}\frac{1+4\frac{\Omega_0^2}{\Omega_1^2}}{(\Gamma_\mathrm{1D}+\Gamma')\Gamma'}\nonumber\\&+\frac{\Gamma'^2+4\Delta_0^2}{4\Gamma_1(\Gamma_\mathrm{1D}+\Gamma')\frac{\Omega_0^2}{\Omega_1^2}}\left[1+\frac{\Gamma_0}{\Gamma'}\left(1+4\frac{\Omega_0^2}{\Omega_1^2}\right)\right].
\label{eq:steadystate}
\end{align}
In Fig.~\hyperref[fig:2]{2(b)}, we compare the analytical prediction from Eq.~(\ref{eq:steadystate}) with the steady-state infidelity obtained from the total system dynamics as a function of $\Gamma_{\mathrm{1D}}/\Gamma'$. We observe near-perfect agreement between the numerical results and our analytical expression.

The steady-state population can be optimized with respect to the ratio of the Rabi frequencies, which we define as
\begin{equation}
\mathcal{R} \equiv \left( \frac{\Omega_0}{\Omega_1} \right)^2\Bigg|_{\mathrm{opt}}.
\label{eq:optim_ratio}
\end{equation}
If we assume $\Gamma_\mathrm{1D}\gg\Gamma'\gg\Delta_i$, the optimal ratio is given by
\begin{equation}
\mathcal{R}=\sqrt{\frac{\Gamma'}{2\Gamma_1}\frac{\Gamma'+\Gamma_0}{4\Gamma'+\Gamma_1}}.
\label{eq:opt_r}
\end{equation}
In the symmetric case $\Gamma_0=\Gamma_1=\Gamma'/2$, we find the optimal value $\Omega_1={3}^{1/4}\,\Omega_0$, and the cooperativity-limited steady-state infidelity becomes
\begin{equation}
(1-F_{T})_{\mathrm{coop}}\approx4.1\frac{\Gamma'}{\Gamma_\mathrm{1D}+\Gamma'}\approx4.1\frac{\Gamma'}{\Gamma_\mathrm{1D}}.
\label{eq:errstat}
\end{equation}
Hence, the final infidelity of our protocol scales as $\sim1/C$, where $C=\beta/(1-\beta)$ and $\beta=\Gamma_\mathrm{1D}/(\Gamma'+\Gamma_\mathrm{1D})$. Such favorable scaling strongly supports the possibility of generating high-fidelity entangled states in realistic systems. In addition, the proportionality constant is sufficiently small that high fidelities can be achieved even for relatively modest coupling strengths. In contrast, the microwave protocol presented in the Appendix, scales with proportionality constant equal to $7$, instead. Moreover, our protocol achieves the steady state with very few requirements, facilitating its experimental implementation. In particular, since the entanglement is reached in steady state, the protocol does not rely on precise timing of experimental pulse sequences.

\subsection{INTRINSIC ERRORS\label{sec:intrinsicerrors}}

Until now, we have studied the system neglecting intrinsic errors that originate from the necessary detunings in the optical pumping processes. On the one hand, as we can see in the effective decay rates presented in the previous section, large detuning implies less efficient pumping into $\ket{T}$, since the fast dynamics from $\gamma_F$ become less efficient. On the other hand, as we will explain in detail below, the protocol requires some detuning to exist between the drivings. This avoids the evolution into separable dark states, which would not be of interest. Here, we study both effects and find a balance between them.

We first study the effect of large detuning. To get the infidelity in Eq.~(\ref{eq:errstat}), we have neglected the contribution from the detunings in Eq.~(\ref{eq:steadystate}). If we include them, we obtain the first intrinsic error for our protocol given by
\begin{gather}
(1-F_{T})_{\mathrm{detu}}=\frac{\Delta_0^2}{\Gamma_\mathrm{1D}\Gamma_1}k_0+\frac{\Delta_1^2}{\Gamma_\mathrm{1D}\Gamma'}k_1,
\label{eq:errdetu}
\end{gather}
which indicates how the infidelity increases with the detunings. Here,
\begin{equation}
k_0 = \frac{\Omega_1^2}{\Omega_0^2}+\frac{\Gamma_0}{\Gamma'}\left(\frac{1+4\frac{\Omega_0^2}{\Omega_1^2}}{\frac{\Omega_0^2}{\Omega_1^2}}\right),
\end{equation}
\begin{equation}
k_1 = 2+8\frac{\Omega_0^2}{\Omega_1^2}.
\end{equation}

As discussed above, some finite detuning is nevertheless necessary. To understand why, consider a single $\Lambda$ system where both ground states are driven to the excited state. The ground-state manifold can then be expressed in terms of a bright state $\ket{b}$ and a dark state $\ket{d}$. The bright state, $\ket{b}\propto(\Omega_0 e^{i\phi_j}\ket{0}+\Omega_1\ket{1})$, is the superposition of ground states that is driven to the excited state $\ket{e}$ and is therefore subject to effective decay. In contrast, the dark state, $\ket{d}\propto(\Omega_1\ket{0}-\Omega_0 e^{i\phi_j}\ket{1})$, does not couple to $\ket{e}$ and is thus not directly driven. However, since the excited state can still decay into both ground states, population will accumulate in $\ket{d}$. To avoid this, the drivings from $\ket{0}$ and $\ket{1}$ have to have different detunings. If the difference in the detunings 
\begin{equation*}
\delta\equiv\Delta_1-\Delta_0
\end{equation*}
is small enough, the population mixing between bright and dark states is too slow, and the system is trapped in $\ket{d}$. Note that, if the total system consists of multiple $\Lambda$-type emitters with similar dynamics, every emitter would independently evolve into a dark-ground state, leading to a separable steady state. Hence, the system requires a sufficiently large $\delta$ to create entangled steady states.

For our protocol, a practical way to study this effect is to distinguish the dark and bright states as those coupled and uncoupled to subradiant states. Therefore, we choose the following basis: \{$\ket{B}=\frac{1}{\sqrt{2\Omega_0^2+\Omega_1^2}}(\sqrt{2}\Omega_0\ket{00}-\Omega_1\ket{S})$, $\ket{D}=\frac{1}{\sqrt{2\Omega_0^2+\Omega_1^2}}(\Omega_1\ket{00}+\sqrt{2}\Omega_0\ket{S})$, $\ket{T}$, $\ket{11}$\} where $\ket{B}$ ($\ket{D}$) is the bright (dark) state coupled to the subradiant (superradiant) manifold. We treat $\ket{11}$ independently.

Following the effective operator formalism presented in the last section, we find the effective dynamics for this basis to be
\begin{equation}
H_{\mathrm{eff}}-H_g\sim\frac{\sqrt{2}\Omega_0\Omega_1\delta}{2\Omega_0^2+\Omega_1^2}(\ketbra{B}{D}+\ketbra{D}{B}),
\end{equation}
\begin{equation}
L_{\mathrm{eff},j}^{(0)}=\sqrt{\frac{\Gamma_0\Omega_0^2}{\Gamma'^2}}\ketbra{B}{B}+\sqrt{\frac{\Gamma_0\Omega_1^2}{2\Gamma'^2}}\ketbra{D}{B},
\end{equation}
\begin{eqnarray}
L_{\mathrm{eff},j}^{(1)}&&=\sqrt{\frac{\Gamma_1}{4}\frac{2\Omega_0^2+\Omega_1^2}{\Gamma'^2}}\ketbra{T}{B}+\sqrt{\frac{\Gamma_1\Omega_0^2}{2\Gamma'^2}}\ketbra{D}{B}\nonumber\\&&-\sqrt{\frac{\Gamma_1\Omega_1^2}{4\Gamma'^2}}\ketbra{B}{B}+\sqrt{\frac{1}{2}\frac{\Gamma_1\Omega_0^2}{\Gamma_\mathrm{1D}^2}}\ketbra{11}{T},\quad\quad
\end{eqnarray}
\begin{equation}
L_{\mathrm{eff}}^{(c)}=\sqrt{\frac{2\Omega_0^2+\Omega_1^2}{2\Gamma_\mathrm{1D}}}\ketbra{D}{T}+\sqrt{\frac{2\Omega_1^2}{\Gamma_\mathrm{1D}}}\ketbra{T}{11}.
\end{equation}
In this case, it is important to consider $H_{\mathrm{eff}}$, as it uniquely describes the pumping out of $\ket{D}$. We incorporate the effective Hamiltonian $H_{\mathrm{eff}}$ together with the effective Lindblad operators $L_{\mathrm{eff}}$ into the master equation and solve for the steady state. We find the infidelity of $\ket{T}$ to yield an additional contribution
\begin{equation}
(1-F_{T})_{\mathrm{dark}} = \frac{\Omega_0^4}{\Gamma_\mathrm{1D}\Gamma'\delta^2}k_2,
\label{eq:errdark}
\end{equation}
which shows that the infidelity increases inversely with the square of the detuning difference, $\delta^2$, and
\begin{equation}
k_2 = \frac{\left(1+2\frac{\Omega_0^2}{\Omega_1^2}\right)^3\left(2\frac{\Gamma_0}{\Gamma_1}+1+4\frac{\Omega_0^2}{\Omega_1^2}\right)}{16\left(\frac{\Omega_0^2}{\Omega_1^2}\right)^3}.
\end{equation}

Equations (\ref{eq:errdetu}) and (\ref{eq:errdark}) describe the intrinsic errors of the protocol. By minimizing the total intrinsic error with respect to the detunings, for a fixed driving strength, we obtain the optimal detunings that minimize the sum of infidelities as
\begin{equation}
\Delta_{0}^{\mathrm{opt}} = -\Omega_0\left({\frac{k_2(k_1\Gamma_1)^3}{k_0(k_0\Gamma'+k_1\Gamma_1)^3}\frac{\Gamma_1}{\Gamma'}}\right)^{1/4},
\label{eq:opt2}
\end{equation}
\begin{equation}
\Delta_{1}^{\mathrm{opt}} = \Omega_0\left({\frac{k_2(k_0\Gamma')^3}{k_1(k_0\Gamma'+k_1\Gamma_1)^3}}\right)^{1/4}.
\label{eq:opt3}
\end{equation}

\begin{figure}[ht]
\centering
    \includegraphics{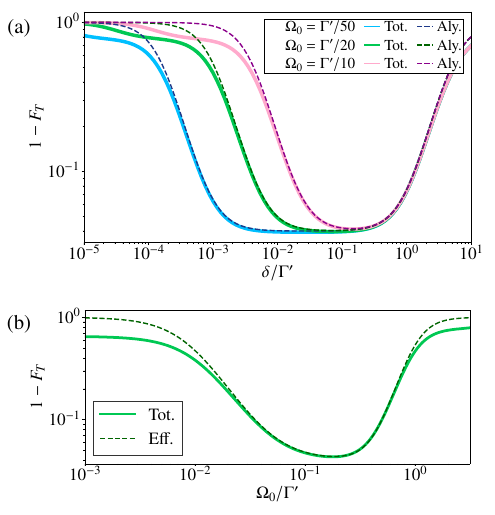}
\caption{\label{fig:3} Comparison of the infidelity for the numerical simulation of the full system (solid) and the analytical results (dashed). (a) Infidelity of the steady state with respect to $\ket{T}$ as a function of the difference in the detunings $\delta = \Delta_1 - \Delta_0$, with $\Delta_1 = -\Delta_0$, for $\Omega_0 / \Gamma' = 0.10$, $0.05$, and $0.02$. Different detuning regimes can be identified. For small $\delta$, the system experiences population trapping in dark states, leading to an infidelity given by $(1-F_{T})_{\mathrm{dark}}$ [Eq.~(\ref{eq:errdark})]. At intermediate $\delta$, a plateau appears where the infidelity is limited by the cooperativity, $(1-F_{T})_{\mathrm{coop}}$ [Eq.~(\ref{eq:errstat})]. For large $\delta$, detuning-induced errors dominate and the infidelity is described by $(1-F_{T})_{\mathrm{detu}}$ [Eq.~(\ref{eq:errdetu})]. We observe a close match between the numerical simulation of the total system dynamics and the analytical results. (b) Infidelity including dephasing. In the presence of ground state dephasing, it is no longer desirable to have a very slow protocol, and $\Omega_0$ can be optimized based on the other parameters. The values used for these plots are $\Gamma_1=\Gamma_2$, $\beta=0.99$, $\Omega_1=\Omega_0\mathcal{R}$, and in panel (b) $\Delta_0=-\Delta_1=\Gamma'/20$ and $\Gamma_\mathrm{d}=10^{-5}\Gamma'$.}
\end{figure}

To verify the analytical predictions, we compare the analytical infidelity $
(1-F_{T})\approx(1-F_{T})_{\mathrm{coop}}+(1-F_{T})_{\mathrm{detu}}+(1-F_{T})_{\mathrm{dark}}
$ with the numerical steady-state infidelity of the entire system in Fig.~\hyperref[fig:3]{3(a)}. As seen in the figure, the two curves are in excellent agreement.

From Eq.~(\ref{eq:errdark}), we observe that small values of $\delta$ reduce the fidelity due to population trapping in the dark state; however, this effect can be mitigated by slowing down the protocol, i.e., by reducing the driving strength $\Omega_0$. Lowering $\Omega_0$ also allows us to decrease $\delta$, thus reducing both intrinsic errors simultaneously while keeping the cooperativity limited error given by Eq.~(\ref{eq:errstat}). This is also shown in Fig.~\hyperref[fig:3]{3(a)} by plotting the infidelity as a function of $\delta$ for different driving strengths.

\subsection{DEPHASING\label{sec:t2}}

In the discussion above, optimal performance is achieved in the limit $\Omega\rightarrow0$, where the protocol is very slow. In practice, this is often not desirable since competing effects may affect the system. To illustrate this, we consider a generic model, where the ground states are subject to dephasing, which drives the system into a mixed state~\cite{Paladino_2014}. This decoherent effect is described by the Lindblad operators $L^{\mathrm{d(0)}}_{j} = \sqrt{\Gamma_{\mathrm{d}}}\ketbra{0}{0}_j$ and $L^{\mathrm{d}(1)}_{j} = \sqrt{\Gamma_{\mathrm{d}}}\ketbra{1}{1}_j$, which effectively create a two-way decay between $\ket{T}$ and $\ket{S}$ at a rate $\Gamma_{\mathrm{d}}$. We can implement this error source into our analysis of the effective dynamics, which results in an additional infidelity contribution given by
\begin{equation}
(1-F_{T})_{\mathrm{deph}}=\frac{\Gamma_\mathrm{d}}{\Omega_0^2}\left[\frac{2\Gamma'\Gamma_1\Omega_0^2/\Omega_1^2+\Gamma_0\Gamma'}{\Gamma_1}\right].
\label{eq:puredeph}
\end{equation}
Note that this error scales inversely with $\Omega_0^2$, which limits how slowly the system can be driven before the state fidelity decreases. In Fig.~\hyperref[fig:3]{3(b)}, we compare the analytical result, adding the last contribution with the total system dynamics. We again obtain a good match between the two descriptions. This additional error gives a lower limit on the speed of the protocol and allows for the optimization of the optical fields' strength.

\section{\label{sec:scaling}SCALING TO MORE EMITTERS}

Above, we have explained and analyzed the scheme for two emitters. However, the scheme directly generalizes to an arbitrary number $N$ of emitters. In the following, we present this generalization and numerically analyze the steady-state fidelity for different $N$. The objective is to create the W-type entangled state
\begin{equation}
\ket{W_N}=\frac{1}{\sqrt{N}}\left(\ket{011...11}+\ket{101...11}+\cdots+\ket{111...10}\right).
\end{equation}
Again, we assume that the decay from $\ket{e}$ to $\ket{0}$ ($\ket{1}$) is (not) coupled to the waveguide. The scheme works the same as the protocol we have presented previously. That is, $\ket{W_N}$ and $\ket{1}^{\otimes N}$ are only coupled to superradiant states. $\ket{1}^{\otimes N}$ decays into $\ket{W_N}$ at a slow rate $\sim\gamma^S$; however, only non-guided decays at rates $\sim\gamma^{ES}$ feed $\ket{1}^{\otimes N}$. The rest of the ground states are coupled to the subradiant manifold, and are either quickly mixed or decay to $\ket{W_N}$ at a rate $\sim\gamma^F$. 

The dynamics for $N$ emitters are qualitatively similar to the two-atom case. The driving described by the Hamiltonian in Eqs.~(\ref{eq:ham1.1})–(\ref{eq:ham1.3}) is essentially the same, except for the phase differences between the optical drivings from $\ket{0}_j$. We choose these phases to maximize the coupling of the ground states to the subradiant manifold. The set of phases that we have observed to achieve maximal coupling is given by
\begin{equation*}
\phi_j=\frac{2\pi(j-1)}{N}, \quad\mathrm{for}\quad 1\le j\le N.
\end{equation*}
The optical pumping from $\ket{1}_j$ might have an emitter dependent phase as well. This leads to different relative phases between the nonlocally superposed states, which can be used for the generation of different W-type ground states.

\begin{figure}[t]
\includegraphics{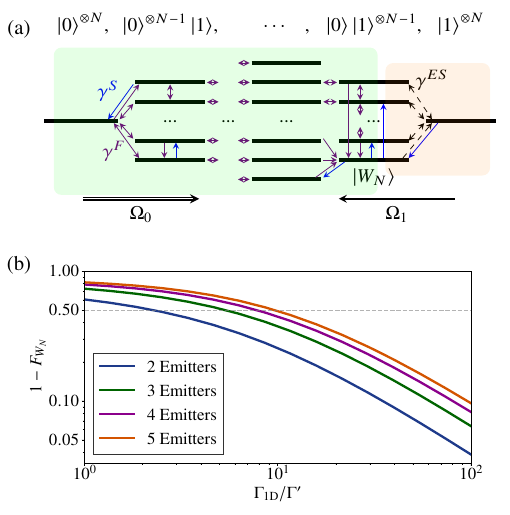}
\caption{(a) Dynamics of the ground states for $N$ $\Lambda$-type emitters. $\ket{W_N}$ is the state we want to prepare. The number of states to the left of $\ket{W_N}$ grows with $N$, but not the number to its right. Increasing $\Omega_0/\Omega_1$ is equivalent to increasing the effective decay rate toward the further-right states. Thus, for optimal results, $\Omega_0/\Omega_1$ should increase with $N$. (b) Steady state infidelity for $\ket{W_N}$ as a function of $C$ for different $N$. For $N=2$, we use the optimal $\Omega_1$, $\Delta_0$, and $\Delta_1$ derived in the text. For $N=3$, $4$, and $5$, we use $\Delta_0=0.4\cdot\Omega_0$ and $\Delta_1=-0.7\cdot\Omega_0$; and the ratio of driving strengths is $N=3$: $\Omega_1=0.88\times\Omega_0$; $N=4$: $\Omega_1=0.69\times\Omega_0$; $N=5$: $\Omega_1=0.59\times\Omega_0$. We have also assumed $\Gamma_0=\Gamma_1=\Gamma'/2$ and $\Omega_0=\Gamma'/100$.\label{fig:4}}
\end{figure}

Increasing the number of emitters also means that the number of ground states increases exponentially. Most of these additional ground states are coupled to the subradiant manifold and are mixed rapidly. For optimal results, we find that decreasing the ratio $\Omega_1/\Omega_0$ with $N$ is beneficial, and helps to drive the population from the rapidly mixed ground states toward the target $\ket{W_N}$ state, thereby improving the steady-state fidelity; see Fig.~\hyperref[fig:4]{4(a)}.

We have studied the scheme for $N=3,4$, and $5$, and the results are presented in Fig.~\hyperref[fig:4]{4(b)}. We have numerically obtained the optimal $\Omega_1/\Omega_0$ to be $\Omega_1/\Omega_0\sim 0.88$ for $N=3$, $\Omega_1/\Omega_0\sim 0.69$ for $N=4$, and $\Omega_1/\Omega_0\sim 0.59$ for $N=5$. In every case, the infidelity scales as $(1-F)\sim 1/C$, and we obtain fidelities above $1/2$ even for relatively low values of $C$. Additionally, we find the steady-state proportionality constants for $(1-F)\sim 1/C$ to increase at a less-than-linear rate.

\section{\label{sec:example}EXPERIMENTAL IMPLEMENTATIONS}

In this section, we analyze the implementation of the scheme for a realistic case of cold trapped $^{133}$Cs atoms. We imagine the emission into the guided mode to dominate over the free space decay by using optical tweezers to trap the atoms in the evanescent field of a bidirectional waveguide. For that, simple one-dimensional dielectric waveguides suffice, provided that sufficiently large $\Gamma_\mathrm{1D}/\Gamma^\prime$ is achieved. As illustrated in Fig.~\hyperref[fig:1]{1(a)}, we envision implementations based on nanophotonic dielectric waveguides, such as photonic-crystal waveguides of the type presented in Ref.~\cite{hood_atomatom_2016}, where the atom coupling to the waveguide is enhanced through slow light effects near a photonic band edge. To achieve a transition-dependent coupling, we consider different hyperfine levels of the atom. Different hyperfine level transitions have different polarizations, which couple differently to the waveguide. The aim of this section is to show that the simplicity of the scheme allows the preparation of entangled states in a realistic experiment. To demonstrate this, we independently consider three types of error. Later, we combine them and show that it is possible to prepare high-fidelity states in the presence of all three errors simultaneously.

We emphasize that the Appendix presents an alternative protocol based on microwave driving between the ground states. While limited to the preparation of two-particle entangled states, this scheme offers a similar fidelity scaling with several theoretical advantages, including the experimental benefits of addressing atomic ensembles with microwave fields (highly homogeneous fields with long wavelengths that are little affected by the presence of, e.g., dielectric waveguide structures). Moreover, the analysis of the different sources of error presented here naturally extends to the protocol discussed there.

\subsection{\label{subsec:4levels}ADDITIONAL GROUND STATES}

\begin{figure}[t]
\includegraphics{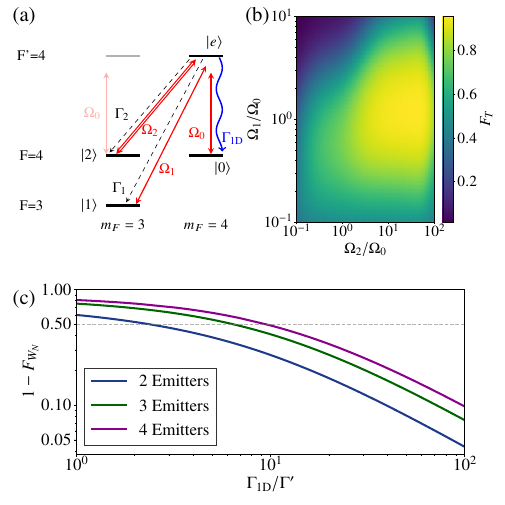}
\caption{\label{fig:5}(a) Considered $^{133}$Cs levels scheme. The transition from $\ket{e}$ to $\ket{0}$ is well coupled to the waveguide, which we assume to be $\pi$ polarized. The entanglement is prepared between the states $\ket{0}$ and $\ket{1}$, and an additional source of error comes from $\ket{e}$ also decaying into $\ket{2}$. Therefore, we introduce a strong driving between $\ket{2}$ and $\ket{e}$ to pump the population out of that state. (b) Steady state fidelity of $\ket{T}$ as a function of the ratios between the drivings for two emitters with $\beta=0.99$. For this simulation, we use the decay rates from the $^{133}$Cs, $F'=4$, and $m_{F'}=4$ level. The parameters used are $\Omega_0=\Gamma'/100$ and $\Delta_0=-\Delta_1=\Omega_2/2=\Gamma'/5$. (c) Steady state fidelity of $\ket{W_N}$ including the additional ground state as a function of $\Gamma_\mathrm{1D}/\Gamma'$ for different $N$. The parameters we use are $N=2$: $\Omega_1= 1.14\times\Omega_0$; $N=3$: $\Omega_1=0.88\times\Omega_0$; $N=4$: $\Omega_1=0.69\times\Omega_0$; and for all of them $\Delta_0=-\Delta_1=\Omega_2/2=\Gamma'/5$ and $\Omega_0=\Gamma'/100$.}
\end{figure}

Using the hyperfine structure, pure $\Lambda$ systems are not always achievable. Typically, the excited state can decay to multiple ground states, which violates the idealized three-level structure. For trapped $^{133}$Cs atoms, a possible implementation of our protocol uses the following levels:
\begin{align*}
&\ket{F=4, m_F=4} = \ket{0}, \\
&\ket{F=3, m_F=3} = \ket{1}, \\
&\ket{F=4, m_F=3} = \ket{2}, \\
&\ket{F'=4, m_{F'}=4} = \ket{e}, 
\label{eq:cg_coef}
\end{align*}
where the ground states $\ket{0}$, $\ket{1}$, and $\ket{2}$ belong to the $6^2S_{1/2}$ manifold, while the excited state $\ket{e}$ is in the $6^2P_{3/2}$ manifold (see Fig.~\hyperref[fig:5]{5(a)}). The additional state $\ket{2}$ is unavoidable due to the hyperfine structure. We choose these $\mathrm{D_2}$ transitions because they provide favorable Clebsch–Gordan coefficients for the individual transitions between $\ket{e}$ and the ground states. These coefficients are given by $
\Gamma_{\ket{e}\leftrightarrow\ket{0}}=\Gamma_0 = \frac{7}{15}\Gamma'$, $
\Gamma_{\ket{e}\leftrightarrow\ket{1}}=\Gamma_1 = \frac{5}{12}\Gamma'$, $
\Gamma_{\ket{e}\leftrightarrow\ket{2}}=\Gamma_2 = \frac{7}{60}\Gamma'$, where $\Gamma'$ is the total decay rate from $\ket{e}$. The only transition with $\pi$ polarized light is between $\ket{e}$ and $\ket{0}$, which we assume to be well coupled to the waveguide. Hence, the other two occur due to non-guided decay.

One of the issues of the additional ground states comes from having multiple transitions not well coupled to the waveguide. Let us illustrate this in the case of two emitters, each with three ground states. The target state is now $\ket{T}\propto(\ket{0}_1\ket{1}_2+\ket{1}_1\ket{0}_2)$. In this case, since both transitions $\ket{1}\leftrightarrow\ket{e}$ and $\ket{2}\leftrightarrow\ket{e}$ are not coupled to the waveguide, there is a symmetric ground state with equivalent characteristics given by $\ket{\Tilde{T}}\propto(\ket{0}_1\ket{2}_2+\ket{2}_1\ket{0}_2)$. To show how to mitigate the effect of $\ket{\Tilde{T}}$, we present the Hamiltonian dynamics with the additional ground state. They are given by
\begin{equation}
H = H_0+V_++V_-,
\label{eq:ham3.1}
\end{equation}
\begin{equation}
H_0 = -\sum_{j=1,2}\left[\Delta_0\ketbra{0}{0}_j+\Delta_1\ketbra{1}{1}_j+\Delta_2\ketbra{2}{2}_j\right],
\label{eq:ham3.2}
\end{equation}
\begin{equation}
V_+ = \sum_{j=1,2}\left[\frac{\Omega_0}{2}e^{i\phi_j}\ketbra{e}{0}_j+\frac{\Omega_1}{2}\ketbra{e}{1}_j+\frac{\Omega_2}{2}\ketbra{e}{2}_j\right],
\label{eq:ham3.3}
\end{equation}
and $V_-=V_+^\dagger$. To avoid population stagnation in $\ket{\Tilde{T}}$, (1) we select $\ket{2}$ such that the nonguided decay into it is the smallest, and (2) we add an additional strong driving field with Rabi frequency $\Omega_2$ to deplete the state. Using the effective operator formalism, we obtain that the population ratio between $\ket{T}$ and $\ket{\Tilde{T}}$ in the steady state is given by
\begin{equation}
{P}_{\Tilde{T}}\approx\frac{\Gamma_2}{\Gamma_1}\frac{\Omega_1^2+2\Omega_0^2}{\Omega_2^2+2\Omega_0^2}F_{T},
\end{equation}
where $F_T=\bra{T}\rho\ket{T}$ and $P_{\Tilde{T}}=\bra{\Tilde{T}}\rho\ket{\Tilde{T}}$. By choosing $\Omega_2\gg\Omega_1, \Omega_0$, we can thus reduce the population $P_{\Tilde{T}} \ll F_T$. In Fig.~\hyperref[fig:5]{5(b)}, we depict an example of the steady-state fidelity of $\ket{T}$ as a function of $\Omega_2/\Omega_0$ and $\Omega_1/\Omega_0$. The limit for how much $\Omega_2$ can be increased is given by the breakdown of the weak coupling regime. We have found numerically that for an optimal fidelity we have to use $\Omega_1/\Omega_0$ dependent on $N$, $\Delta_0=-\Delta_1=\Omega_2/2=\Gamma'/5$, and $\Delta_2=0$. A high $\Omega_2/\Omega_0$ has the additional advantage that it increases the probability of keeping the electron within the states of interest.

Thus, we have identified a way to minimize the impact of the additional ground states on the steady-state fidelity of the protocol. We can now extend this approach to a larger number of atoms. The target state remains $\ket{W_N}$, and it is optimally prepared by choosing the phases of the optical drivings from $\ket{0}_j$ as described in the previous section. Figure~\hyperref[fig:5]{5(c)} shows the steady-state fidelity for $N = 2$, $3$, and $4$ emitters. We numerically find an optimal ratio between the driving for $N=2$: $\Omega_1= 1.14\times\Omega_0$; $N=3$: $\Omega_1=0.88\times\Omega_0$; $N=4$: $\Omega_1=0.69\times\Omega_0$; and for all of them we choose $\Delta_0=-\Delta_1=\Omega_2/2=\Gamma'/5$ for $\Omega_0=\Gamma'/100$. We observe that the protocol reliably drives the system into high-fidelity entangled states, with similar scaling as the case of perfect $\Lambda$-type emitters with $(1-F_{T})\sim 1/C$.

\subsection{\label{subsec:motion}EFFECTS OF ATOMIC MOTION}

Another experimental imperfection arises from the motion of the emitters. Our scheme relies on the spacing between atoms resulting in superradiant and subradiant states, and having no dispersive interaction, which is altered if the atoms are perturbed from their ideal rest positions. Hence, we want to study the effect of atomic motion on the state-preparation fidelity. 

To model this error, we consider the atoms trapped in harmonic oscillator potential wells. We study the motion along the waveguide's axis, since movement in this direction has the largest disturbance on the subradiant and superradiant states. The Hamiltonian that describes the motion of the emitters is given by
\begin{equation}
H_{M}=\sum_j\omega_z a^\dagger_ja_j,
\end{equation}
where $\omega_z$ is the frequency of the trapped atoms along the waveguide axis, and $a_j$ ($a^\dagger_j$) is the annihilation (creation) operator for phonons on the $j$th atom. Furthermore, driving transitions on the internal states of the atom also induce couplings between different motional states~\cite{wineland1998}. The coupling is given by
\begin{eqnarray}
\Tilde{V} =& \sum_j\Bigl[\frac{\Omega_0}{2}e^{i\phi_j}\ketbra{e}{0}_j e^{i\eta(a_j+a_j^\dagger)}\nonumber \\&+ \frac{\Omega_1}{2}\ketbra{e}{1}_j e^{i\eta(a_j+a_j^\dagger)}+\mathrm{H.c.}\Big].
\end{eqnarray}
where $\eta= k\,z_0$ is the Lamb-Dicke parameter, $k=2\pi/\lambda$ is the wavenumber of the incident fields, $z_0=(\hbar/2m\omega_z)^{1/2}$ is the spread of the zero-point wavefunction and $m$ the mass of the atom. At the same time, decay can also induce transitions between internal and motional states. The Lindblad operators are hence modified to
\begin{equation}
\Tilde{L}_{j}^{(0)} = \sqrt{\Gamma_0}\ketbra{0}{e}_j e^{-i\eta(a_j+a_j^\dagger)/\sqrt{3}},
\end{equation}
\begin{equation}
\Tilde{L}_{j}^{(1)} = \sqrt{\Gamma_1}\ketbra{1}{e}_j e^{-i\eta(a_j+a_j^\dagger)/\sqrt{3}},
\end{equation}
\begin{equation}
\Tilde{L}_{\pm}^{(c)} = \sqrt{\frac{\Gamma_\mathrm{1D}}{2}}\left[\ketbra{0}{e}_1 e^{\pm i\eta(a_1+a_1^\dagger)}+\ketbra{0}{e}_2 e^{\pm i\eta(a_2+a_2^\dagger)}\right].
\end{equation}
We have here added a heuristic geometric factor of $1/\sqrt{3}$ to the exponents of the single emitter decays that represents the fraction of the total phonon recoil that contributes to the chosen axis motion. This is not the case for the collective decay since it is completely along the waveguide's axis. Moreover, the collective decay has to be split into two to distinguish between right and left propagating photons.

\begin{figure}[t]
    \includegraphics{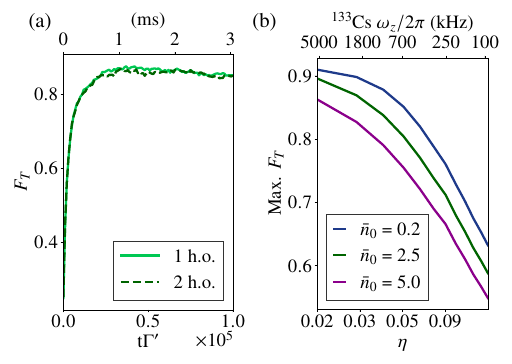}
\caption{\label{fig:6} Fidelity of the entangled state $\ket{T}$ considering two emitters trapped in harmonic oscillator wells. (a) Monte Carlo simulation for the fidelity of $\ket{T}$ considering both relative and the center-of-mass motion (2 h.o.) and only the relative motion (1 h.o.). The atoms start from a completely mixed state and the phonons in independent thermal states with mean phonon number $\bar{n}(0)$. Since we do not include any cooling procedure, we see the fidelity reaching a maximum and subsequently decaying due to the emitters heating up. For this simulation we have chosen a cut-off of the vibrational state at $N_c=20$, $\Omega_0=\Gamma'/20$, $\bar{n}_0=0.2$ and $\eta=0.05$, which corresponds to a trapping frequency along the waveguide's axis of $\omega_z/2\pi\approx0.84$ MHz for $^{133}$Cs atoms. (b) Maximum of the fidelity as a function of $\eta$ and its corresponding $\omega_z$ for $^{133}$Cs, for $\bar{n}(0)=0.2,\,2.5,$ and $5.0$. We consider $N_c=15$ and $\Omega_0=\Gamma'/30$. We observe how minimizing $\eta$ and $\bar{n}(0)$ results in increasing the maximum fidelity for the protocol. Other parameters used for the two plots are $\Gamma_0=\Gamma_1$, $\beta=0.98$, and the optimal $\Omega_1$, $\Delta_0$, and $\Delta_1$ derived for the case of two emitters. }
\end{figure}

For simulation purposes, instead of working in the tensor product of the individual motional Hilbert spaces of each emitter, we transform to collective motional modes defined as
\begin{equation*}
a_\mathrm{R}\equiv\frac{a_1-a_2}{\sqrt{2}}\quad\mathrm{and}\quad a_\mathrm{CM}\equiv\frac{a_1+a_2}{\sqrt{2}},
\end{equation*}
where $a_\mathrm{R}$ describes the relative motion between the atoms and $a_\mathrm{CM}$ the center-of-mass motion. For sufficiently small values of $\eta^2 \bar{n}$, we assume that the center-of-mass mode does not significantly affect the system fidelity and therefore neglect it in the simulations. To justify this approximation, Fig.~\hyperref[fig:6]{6(a)} shows the time evolution of the fidelity $F_T$ obtained by including both motional modes, compared with the result obtained by considering only the relative mode. The agreement observed between the two approaches validates this simplification for the chosen parameters.

Since no cooling mechanism is included in the simulations, the atoms are expected to heat up over time. As a consequence, the motional state progressively deviates from the ideal subradiant and superradiant configurations, leading to a gradual reduction of the fidelity at long times, as shown in Fig.~\hyperref[fig:6]{6(a)}. Therefore, the fidelity reaches a maximum at an intermediate time. The maximal fidelity depends on both the Lamb-Dicke parameter $\eta$ and the initial thermal occupation $\bar{n}(0)$. After each optical transition, the probability of changing the motional state scales as $\eta^2$. Thus, larger values of $\eta$ lead to faster heating, lower peak fidelities of the target state, and shorter optimal preparation times. Similarly, a larger initial occupation $\bar{n}(0)$ corresponds to starting from a less favorable motional configuration, further reducing the achievable fidelity.

Figure~\hyperref[fig:6]{6(b)} shows the maximal fidelity obtained for $\ket{T}$ as a function of $\bar{n}(0)$ and $\eta$. For the case of $^{133}$Cs atoms, we estimate $\eta \approx \sqrt{4\times 10^{-4} \Gamma'/\omega_z}$. As shown in the figure, even in the presence of motional effects, the protocol is capable of preparing high-fidelity entangled states at intermediate times. For realistic parameters, the achievable fidelities remain well above the classical threshold $F_T = 1/2$.

We note that for free-space optical tweezers and Cs atoms, typical vibrational frequencies for micron-scale beam waists are on the order of a few hundred kHz. However, higher spatial confinement (MHz) could be achieved by exploiting the strong field gradients of guided-mode optical lattices in nanophotonic waveguides or by trapping in near-field patterns generated by a side-illuminating tweezer focused onto subwavelength dielectric structures, e.g., holes. Far off-resonant scattering from the trapping fields can also contribute to motional heating as well as depumping, however these rates are typically negligible on the timescales of the protocol.

\subsection{\label{subsec:correlated} BROADENING OF TRANSITIONS}

Until now, we have assumed that the driving phase between the emitters is the only asymmetry in the scheme. However, frequency shifts can arise, for example, from differential AC Stark shifts induced by the trapping fields and van der Waals interactions near dielectric surfaces. The former can be minimized by the use of magic wavelengths \cite{katori_2003} and appropriate polarization.
In this section, we discuss the influence of broadening of the transitions. For that, we study two identical atoms and consider an independent variation in their transition detunings. 

We define the ideal detuning $\Delta_{i}$ for the transition $i$ to be equal between both emitters, and introduce a small variation $\Delta_{i} + \epsilon_{ij}$ on emitter $j$. We can separate the effect into 
\begin{align*}
&\sigma^\mathrm{sum}_+=\sum_{i,j} \epsilon_{ij}/4, &\sigma^\mathrm{dif}_+=\sum_{i,j} (-1)^j\epsilon_{ij}/4, \\&\sigma^\mathrm{sum}_-=\sum_{i,j} (-1)^i\epsilon_{ij}/4, & \sigma^\mathrm{dif}_-=\sum_{i,j} (-1)^{j+i}\epsilon_{ij}/4,
\end{align*} where $\sigma^\mathrm{sum}_+$ and $\sigma^\mathrm{dif}_+$ result in optical broadening, and $\sigma^\mathrm{sum}_-$ and $\sigma^\mathrm{dif}_-$ represent broadening for the ground-state transitions. $\sigma^\mathrm{dif}_-$ effectively induces a driving between $\ket{T}$ and $\ket{S}$ and is the broadening with the greatest influence on the steady state. 

\begin{figure}[t]
\resizebox{0.45\textwidth}{!}{
    \includegraphics{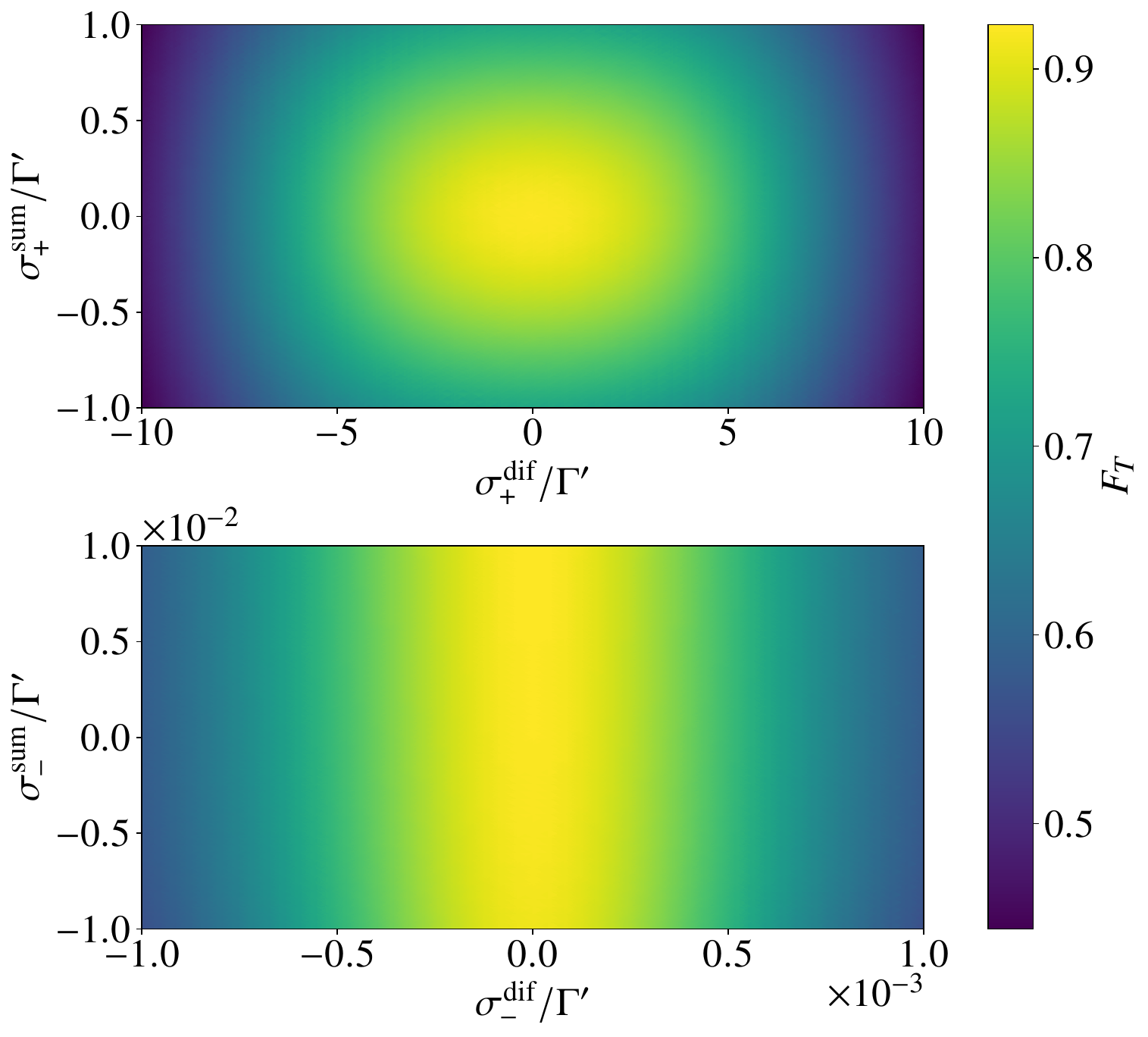}}
\caption{\label{fig:7} Effect of broadening of transitions on the final fidelity of $\ket{T}$. Effect of the broadening of the optical transitions (top) and of the ground-states transitions (bottom). Note that the scheme is most susceptible to $\sigma^\mathrm{dif}_-$, which corresponds to uncorrelated noise between the emitters. In both simulations, we have used $\Gamma_1=\Gamma_2$, $\beta=0.98$ $\Omega_0=\Gamma'/15$, $\Delta_0=-\Delta_1=\Gamma'/50$, and $\Omega_1=\Omega_0\mathcal{R}$.}
\end{figure}

We numerically study the effect by considering two $\Lambda$-type atoms with a wide range of detunings and present it in Fig.~\ref{fig:7}. We observe that for $\beta=0.98$ and $\Omega_0=\Gamma'/15$, with broadenings given by $\sigma^\mathrm{dif}_-/\Gamma'\sim10^{-4}$, we can still get fidelities of $F\gtrsim 0.90$. In the case of the transitions studied for $^{133}$Cs atoms, for $\Gamma'=2\pi\times 5.234$ MHz~\cite{steck_2025}, that is, $T_2^*=2/\sigma^\mathrm{dif}_-\sim 600\mu s$.

\subsection{SUMMARY\label{sec:summaryexample}}

\begin{figure}[t]
\resizebox{0.45\textwidth}{!}{
\includegraphics{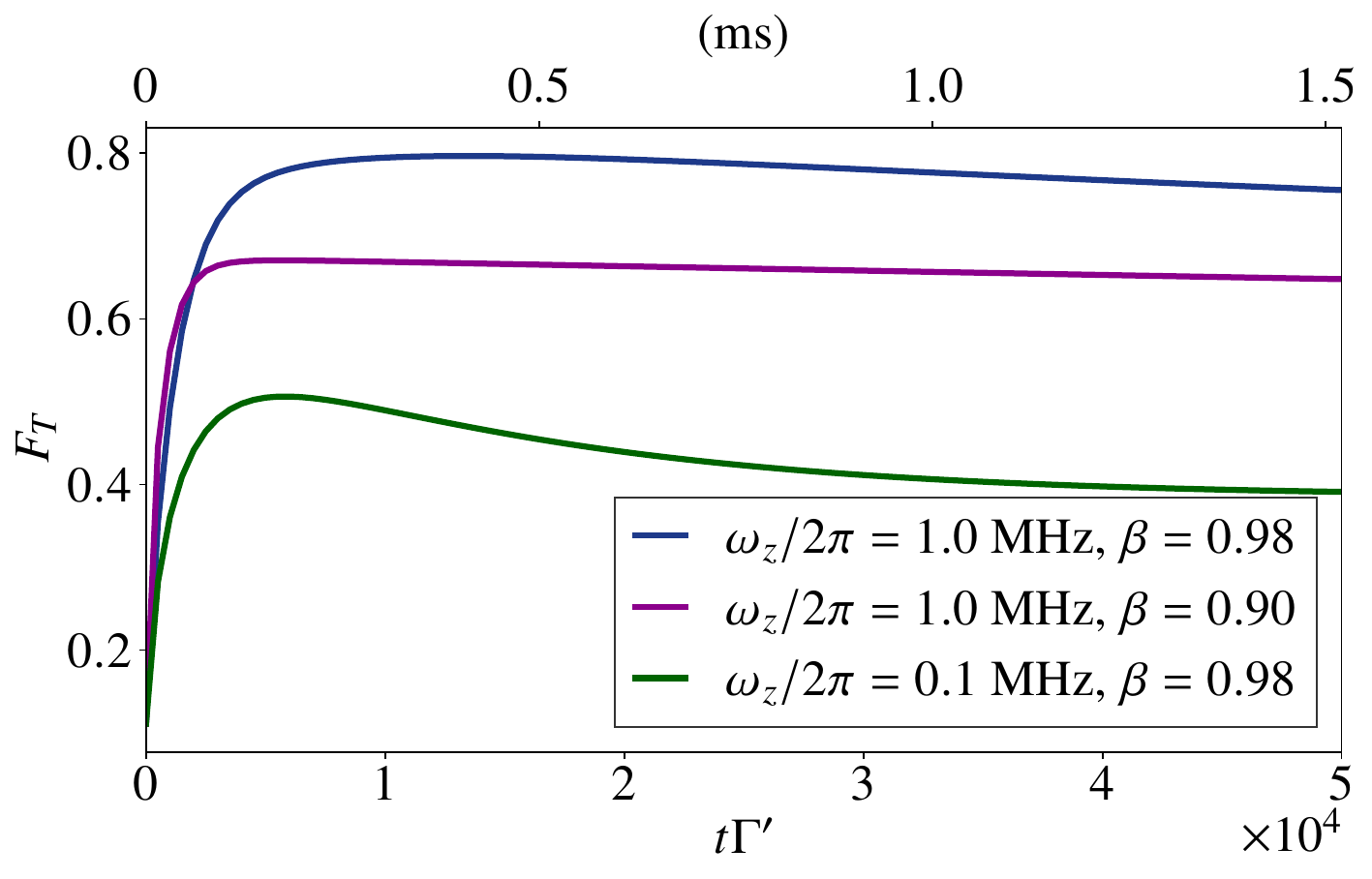}}
\caption{\label{fig:8} Transient fidelity for the $\ket{T}$ state combining all the considered errors. We consider two $^{133}$Cs atoms trapped in harmonic-oscillator-type wells. As the excited state we consider the hyperfine level $6^2P_{3/2}, F'=4, m_{F'}=4$, with a total free space decay rate $\Gamma'$. We study the evolution for trapping frequencies $\omega_z/2\pi\in$ \{1 MHz, 100 kHz\} and coupling $\beta\in$ \{0.90, 0.98\}. We also consider $\Omega_0=\Gamma'/20$, $\Omega_1=\Omega_0\mathcal{R}$, $\Delta_0=-\Delta_1=\Omega_2/2=\Gamma'/5$, $N_c=15$, and $\sigma^\mathrm{dif}_-=10^{-4}\Gamma'$ that corresponds to $T_2^*\sim 600\mu s$, and a relation between the trapping frequency and the Lamb-Dicke parameter of $\eta=\sqrt{4\times 10^{-4}\Gamma'/\omega_z}$. We get the maximum of fidelity $F_T=0.80$ for $\omega_z/2\pi=1$ MHz and coupling efficiency $\beta=0.98$.}
\end{figure}

In this section, we have considered some of the most important experimental issues separately, and we have shown for each case that high fidelities are achievable with realistic experimental parameters. Here, we combine all these potential errors to get the expected realistic behavior. We consider two $^{133}$Cs atoms, including the extra level $6^2P_{3/2}, F'=4, m_{F'}=4$, which decays back to $\ket{2}$; see Fig.~\ref{fig:5}. We include $\Omega_0=\Gamma'/20$, $\Omega_1=\Omega_0\mathcal{R}$, $\Delta_0=-\Delta_1=\Omega_2/2=\Gamma'/5$, and $\sigma^\mathrm{dif}_-=10^{-4}\Gamma'$. We study the system for different values of the coupling efficiency and show the results in Fig.~\ref{fig:8}. For the different parameters, we obtain
\begin{align*}
\omega_z/2\pi=\mathrm{1.0\,MHz},\,\beta=0.98\rightarrow F_\mathrm{max}=0.80, \\
\omega_z/2\pi=\mathrm{1.0\,MHz},\,\beta=0.90\rightarrow F_\mathrm{max}=0.67, \\
\omega_z/2\pi=\mathrm{0.1\,MHz},\,\beta=0.98\rightarrow F_\mathrm{max}=0.51.
\label{eq:realistic_fidelities}
\end{align*}
Hence we observe that high fidelities of the desired state can be obtained for realistic experimental parameters.

\section{CONCLUSIONS\label{sec:conclusions}}

We have proposed a simple protocol for generating high-fidelity multiparticle entangled steady states between the ground levels of $\Lambda$-type emitters coupled to a waveguide. We have shown that the fidelity of the protocol exhibits favorable scaling with both the cooperativity, $(1-F) \sim 1/C$, and the number of emitters.

We have studied the intrinsic limits of the protocol and evaluated the effects of several imperfections relevant to its experimental implementation, including pure dephasing, additional atomic levels, atomic motion, and broadening. Notably, we have shown that, when atomic motion is taken into account, the highest entangled-state fidelity is reached at an intermediate time during the transient evolution.

For a particular implementation based on Cs atoms, we have given a full picture of our protocol with realistic experimental values. We have seen that fidelities well above the classical limit $F=1/2$ can be obtained, even when combining all imperfections. 

Beyond its fundamental interest, dissipative protocols such as the one presented here are particularly attractive experimentally because of their favorable scaling and absence of requirements for precise timing or feedback control mechanisms. While the presented protocol only realizes a particular state, the platform is being explored for more advanced applications. This protocol could be used to benchmark waveguide QED platforms, and as an important precursor for more complex applications. 

\begin{acknowledgments}

We are grateful for financial support from Danmarks Grundforskningsfond (DNRF 139, Hy-Q Center for Hybrid Quantum Networks) and the Novo Nordisk Foundation through the NovoNordisk Center for Biomedical Quantum Sensing (Grant number: NNF24SA0088433). J.A. acknowledges support from the Novo Nordisk Foundation (Challenge project “Solid-Q”). J.H. and J.B. also acknowledge support from the Novo Nordisk Foundation (Grant number: NNF20OC0059939 ‘Quantum for Life’).

\end{acknowledgments}

\section{Data Availability}

The data that support the findings of this article are openly available~\cite{alba2026dataset}.

\appendix
\section{\label{appen:microw}MICROWAVE-DRIVEN PROTOCOL}

\begin{figure*}[ht]
\includegraphics{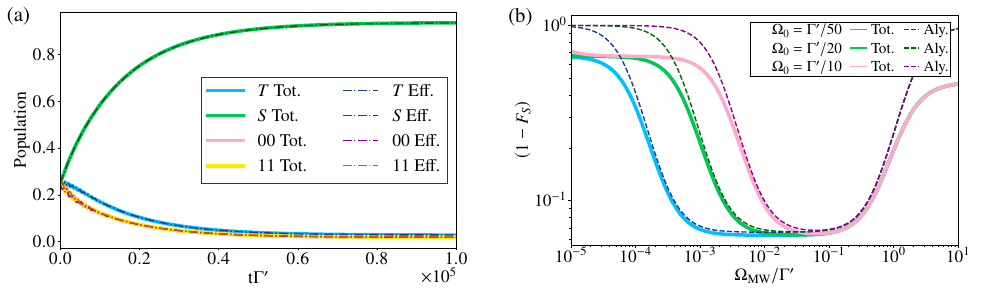}
\caption{\label{fig:a1} Comparison of the total system dynamics, effective dynamics, and analytical results for the microwave-driven protocol. (a) Transient evolution comparing total system (solid) with effective (dashed) dynamics in the singlet-triplet basis. Starting from a completely mixed state, we obtain a near perfect match between the two approaches. We use $\Omega_0=\Gamma'/50$ and $\Omega_\mathrm{MW}=\Gamma'/300$. (b) Infidelity of the steady state with respect to $\ket{S}$ as a function of $\Omega_\mathrm{MW}$. We plot the numerical results (solid) and the analytical errors (dashed), including the contributions from dressing and poor mixing of ground states. We see that, for different values of $\Omega_0$, the numerical results match the analytical errors for the regime with greater fidelity. In both plots, we have used $\beta=0.99$ and $\Gamma_0=\Gamma_1$.}
\end{figure*}

In the following, we present an alternative protocol that substitutes one of the optical drivings for microwave transitions between the ground states. This protocol cannot easily be generalized to more than two emitters, but offers a similar fidelity as the scheme already presented. Similarly to the case we have presented, the protocol reliably drives the system to a high-fidelity steady state without the need for feedback control. The protocol was presented in Ref.~\cite{Gullans_2012}, and here we study it further to give a comparison with the protocol from the main text. 

We study the emitters as $\Lambda$-like systems, with $\ket{0}$ and $\ket{1}$ being the ground states and $\ket{e}$ the excited states. The coherent dynamics in this case are given by
\begin{equation}
H = H_0+V+H_e,
\label{eq:ham2.1}
\end{equation}
\begin{equation}
H_0 = \Delta_\mathrm{MW}\sum_j\ketbra{0}{0}_j+
\frac{\Omega_\mathrm{MW}}{2}\sum_j\left[\ketbra{0}{1}_j+\mathrm{H.c.}\right],
\label{eq:ham2.2}
\end{equation}
\begin{equation}
V =\frac{\Omega_0}{2} \sum_j\left[e^{i\phi_j}\ketbra{0}{e}_j+\mathrm{H.c.}\right],
\label{eq:ham2.3}
\end{equation}
\begin{equation}
H_e = \Delta_\mathrm{0}\sum_j\ketbra{e}{e}_j,
\label{eq:ham2.4}
\end{equation}
where the coupling between ground and excited states is assumed to be weak, $\Omega_0\ll\Gamma'$, where $\Gamma'$ is the total nonguided decay of the excited state. Again, we are interested in creating perfect superradiant and subradiant states, which give the same non-unitary dynamics as those described by the Lindblad operators in Eqs.~(\ref{eq:lind1.1})-(\ref{eq:lind1.3}). 

In this protocol, the desired state is the antisymmetric state $\ket{S}=\frac{1}{\sqrt{2}}(\ket{0}_1\ket{1}_2-\ket{1}_1\ket{0}_2)$. The microwave field mixes the population between the states $\ket{00} \leftrightarrow \ket{T} \leftrightarrow \ket{11}$, and the state $\ket{00}$ is pumped to the subradiant state, which partially decays to $\ket{S}$. This transition is fast since the intermediate excited state decays slowly. However, $\ket{S}$ does not mix with the rest of the ground states and is coupled to a superradiant state. Hence, the pumping out of $\ket{S}$ is slow since the mediating state decays rapidly.

Assuming the optimal fidelities are obtained with $\Delta_\mathrm{MW}=\Delta_0=0$, the eigenstates for the ground-state Hamiltonian $H_0$ are: $\{\ket{S}$, $\ket{U_0} =\frac{1}{\sqrt{2}}\left(\ket{00}-\ket{11}\right)$, $\ket{U_{\pm}}=\frac{1}{2}\left[\ket{00}+\ket{11}\pm \sqrt{2}\ket{T}\right]\}$, which yield the following effective Lindblad operators
\begin{widetext}
\begin{equation}
L_{\mathrm{eff},j}^{(0)}=\sqrt{ \gamma^{FM}_{0}/8}\Bigl[\ket{U_0}\left(\bra{U_+}+\bra{U_-}\right)+\frac{1}{\sqrt{2}}\Bigl(\ketbra{U_-}{U_+}\\+\ketbra{U_+}{U_-}\Bigr)\Bigr]+\sqrt{\gamma^{F}_{0}/8}\Big(\ket{U_+}+\ket{U_-}\Big)\bra{U_0},
\end{equation}
\begin{equation}
L_{\mathrm{eff},j}^{(1)}=\sqrt{\gamma^{F}_{1}/4}\left(-\ket{S}+(-1)^j\frac{1}{\sqrt{2}}\ket{U_+}-(-1)^j\frac{1}{\sqrt{2}}\ket{U_-}\right)\bra{U_0}-(-1)^j\sqrt{ \gamma^{FM}_{1}/8} \Bigl[\ket{S}\left(\bra{U_+}+\bra{U_-}\right)+\frac{1}{\sqrt{2}}\left(\ketbra{U_-}{U_+}+\ketbra{U_+}{U_-}\right)\Bigr],
\end{equation}
\begin{equation}
L_{\mathrm{eff}}^{(c)}=\sqrt{ \gamma^{S}/2}\left(\ket{U_+}+\ket{U_-}\right)\bra{S},
\end{equation}
\end{widetext}
where the different effective decay rates are given by
\begin{equation*}
\gamma^{F}_{m}=\frac{\Gamma_m\Omega_0^2}{\Gamma'^2},\quad\quad \gamma^{FM}_{m}=\frac{\Gamma_m\Omega_0^2}{4\Omega_\mathrm{MW}^2+\Gamma'^2},\quad\quad\gamma^{S}=\frac{\Omega_0^2}{\Gamma_\mathrm{1D}}.
\end{equation*}
If we assume $\Omega_\mathrm{MW}\gg\Omega_0^2/\Gamma'$, then $H_{\mathrm{eff}}\approx H_0$, and the eigenstates of $H_\mathrm{eff}$ are approximate eigenstates of $H_{0}$. This means that the effective drivings between ground states are negligible compared to the effective decays, and the rate equations simplify. Assuming the symmetric case $\Gamma_0=\Gamma_1=\Gamma'/2$, we obtain the steady-state fidelity for $\ket{S}$ to be
\begin{equation}
(1-F_S)_\mathrm{coop} = \frac{7\Gamma'}{\Gamma_\mathrm{1D}}.
\end{equation}
Hence, this protocol also allows for high steady-state fidelities with an error decreasing as $\sim1/C$, although with a slightly less favorable coefficient. This second protocol has some advantages. For example, all the eigenstates of $H_0$ are effectively coupled through the subradiant state to $\ket{S}$. This allows for a faster preparation of the steady state.

\subsection{INTRINSIC ERRORS}

This protocol relies on $\Omega_\mathrm{MW}$ being within an ideal regime. On the one hand, increasing $\Omega_\mathrm{MW}$ too much dresses the ground states $\ket{U_{\pm}}$ and makes it harder to pump out of them. The error added from this contribution increases as
\begin{equation}
(1-F_S)_\mathrm{dre}=\frac{24\Omega_{\mathrm{MW}}^2}{\Gamma'\Gamma_\mathrm{1D}}.
\end{equation}
On the other hand, too little $\Omega_\mathrm{MW}$ slows the mixing between ground states and mainly causes stagnation of the system in $\ket{11}$, which adds the error
\begin{equation}
(1-F_S)_\mathrm{mix} = \frac{4\Omega_0^4}{\Omega_{\mathrm{MW}}^2\Gamma'\Gamma_\mathrm{1D}}.
\end{equation}

In Fig.~\ref{fig:a1}, we depict these results and compare the total system dynamics with the analytical result.

\end{document}

%% file: 1packages.tex
\usepackage{graphicx}
\usepackage{color}
\usepackage[colorlinks=true,citecolor=blue,linkcolor=red,urlcolor=blue]{hyperref}
\usepackage{orcidlink}
\usepackage{amsmath}
\usepackage{amsthm}
\usepackage{physics}
\usepackage{amsfonts}
\usepackage{newtxtext,newtxmath}
\usepackage{cancel}

\usepackage[normalem]{ulem}
\usepackage[final]{changes}

\usepackage{tikz} 
\usetikzlibrary{decorations.markings} 
\usetikzlibrary{shapes.misc}
\usepackage{dcolumn}
\usepackage{bm}

%% file: 2commands.tex
\def\bra#1{\mathinner{\langle{#1}|}}
\def\ket#1{\mathinner{|{#1}\rangle}}

\definecolor{amethyst}{rgb}{0.6,0.4,0.8}
\definecolor{almond}{rgb}{0.94,0.87,0.8}
\definecolor{airforceblue}{rgb}{0.36,0.54,0.66}
\definecolor{aqua}{rgb}{0.0,0.7,1}
\definecolor{darkblue}{HTML}{0317fc}
\definecolor{lightblue}{HTML}{03cffc}
\definecolor{lightgrey}{HTML}{c4c4c4}
\definecolor{dorange}{HTML}{cc8f00}
\definecolor{dred}{HTML}{cf0617}
\definecolor{lorange}{HTML}{ffc400}
\definecolor{lpurple}{HTML}{ff00ae}
\definecolor{dpurple}{HTML}{730083}
\definecolor{turquoise}{HTML}{00e1d7}
\newcommand{\beq}{\begin{equation}}
\newcommand{\eeq}{\end{equation}}

\definecolor{viridis0}{rgb}{0.267004, 0.004874, 0.329415}
\definecolor{viridis1}{rgb}{0.281412, 0.125229, 0.443197}
\definecolor{viridis2}{rgb}{0.232142, 0.353289, 0.527186}
\definecolor{viridis3}{rgb}{0.165344, 0.561481, 0.619823}
\definecolor{viridis4}{rgb}{0.126438, 0.749294, 0.776468}
\definecolor{viridis5}{rgb}{0.033421, 0.898247, 0.996402}

\definecolor{deep1}{HTML}{1E3A8A}
\definecolor{deep2}{HTML}{006400}
\definecolor{deep3}{HTML}{8B008B}
\definecolor{deep4}{HTML}{D35400}

\definecolor{light1}{HTML}{00BFFF}
\definecolor{light2}{HTML}{00C853}
\definecolor{light3}{HTML}{FFAACC}
\definecolor{light4}{HTML}{FFEA00}

\definecolor{darkblue}{HTML}{0317fc}

%% file: 3title.tex

\title{Steady-State Multiparticle Entanglement via Dissipative Engineering \\ in Waveguide QED}

\author{Joan Alba} %
\email{joan.pares@nbi.ku.dk} 
\affiliation{Center for Hybrid Quantum Networks (Hy-Q), Niels Bohr Institute, University of Copenhagen, Jagtvej 155A, 2200 Copenhagen, Denmark}

\author{Jacob Thornfeldt Hansen} 
\affiliation{QUANTOP, Niels Bohr Institute, University of Copenhagen, Blegdamsvej 17, 2100 Copenhagen, Denmark}

\author{Jean-Baptiste S. Béguin} 
\affiliation{QUANTOP, Niels Bohr Institute, University of Copenhagen, Blegdamsvej 17, 2100 Copenhagen, Denmark}

\author{Anders S. Sørensen} 
\affiliation{Center for Hybrid Quantum Networks (Hy-Q), Niels Bohr Institute, University of Copenhagen, Jagtvej 155A, 2200 Copenhagen, Denmark}

\date{\today}
             
\begin{abstract}
We propose a simple scalable scheme for the dissipative generation of entangled states of multiple emitters coupled to a waveguide. Our approach exploits collective interactions arising from the formation of subradiant and superradiant excited states, combined with the quantum Zeno effect. We show that, starting from an arbitrary initial state, the system deterministically evolves toward a W-type entangled steady state, with an infidelity that scales inversely with the cooperativity. The protocol is scalable to an arbitrary number of emitters. We further analyze the impact of additional experimental imperfections and present a detailed implementation based on trapped $^{133}$Cs atoms.
\end{abstract}